\documentclass[12pt]{article}

\usepackage[dvipdfm]{graphicx}
\usepackage{subfigure}
\begin{document}

\title{ Two kinds of Phase transitions in  a Voting model}

\author{M Hisakado\footnote{[1]
masato\_hisakado@standardandpoors.com}  \space{} and   S Mori\footnote{[2] mori@sci.kitasato-u.ac.jp} }

\maketitle

*Standard  \& Poor's, Marunouchi 1-6-5, Chiyoda-ku, Tokyo 100-0005, Japan

\vspace*{1cm}

\dag Department of Physics, School of Science,
Kitasato University, Kitasato 1-15-1, Sagamihara, Kanagawa 252-0373, Japan

 \vspace*{1cm}

\begin{abstract}
In this paper, we discuss a voting model with two candidates, $C_0$ and  $C_1$.
We consider  two types of voters--herders and independents.
The voting of independents  is  based on  their fundamental values;
on the other hand, 
the   voting of herders is  based on the number of  previous votes.
We can identify two kinds of phase transitions.
One is  an information cascade transition similar to a  phase transition   seen in Ising model.
The other is  a transition of super and normal diffusions.
These phase transitions coexist.
We compared our  results to the conclusions of experiments
and identified  the phase transitions in  the upper   limit of the time $t$  by  using analysis  of  human behavior obtained from experiments.

\end{abstract}



\newpage
\section{Introduction}

In general, collective  herding behavior  poses  interesting problems in several cross fields such as sociology \cite{tarde}, social psychology \cite{mil},  ethnology \cite{fish}\cite{frank},  and economics. 
Statistical physics offers effective  tools to analyze these phenomena caused by collective  herding behaviors, and  the associated  field is  known as sociophysics \cite{galam}.  For example, in statistical  physics, anomalous fluctuations in   financial markets \cite{Cont}\cite{Egu} and  opinion dynamics \cite{Stau}\cite{Curty}\cite{nuno} have been related  to percolation and the  random
 field Ising model.

To estimate public perception,  people observe the actions of other individuals; then, they  make a  choice similar to  that of  others. 
 Because it is usually sensible to do what other people are doing, collective  herding behavior  is assumed to be the result of  a rational choice. This approach can sometimes  lead to arbitrary or even erroneous decisions as  a macro phenomenon. This phenomenon  is known as an  information cascade \cite{Bikhchandani}


In our previous paper, we introduced a sequential voting model that is similar to  a Keynesian beauty contest  \cite{Keynes}\cite{Hisakado2}\cite{Mori2}.
At each time step $t$, one  voter  votes   for one  of two  candidates.
As public perception, the  $t$th voter can see all previous votes, i.e. $(t-1)$ votes.
There are two types of voters--herders and independents--and two candidates.
Herders are also  known  as copycat voters; 
they vote for  each candidate with  probabilities that are proportional  to the candidates' votes.
We refer to these herders  as  analog herders.
We investigated a case wherein  all the  voters were herders \cite{Mori}.
In such  a case, the voting  process is a P\'{o}lya process, and the voting rate converges to a beta distribution in a large time  limit of $t$  \cite{Hisakado}.

Next, we added independents to  the  analog herders \cite{Hisakado2}.
In the upper limit of $t$, the  independents cause the distribution of votes to converge to a  Dirac measure against herders. 
This model  contains three phases--two super diffusion phases and a normal diffusion phase.
We refer to  the transition in this model  as a transition of   super and normal diffusions. 
These transitions  can be seen in several  fields \cite{Hod}.
If herders constitute  the   majority or even  half of the total voters,
the voting rate  converges  to  a Dirac measure    slower   than   in a  binomial distribution.
These  two phases  have    different speeds of convergence that are    slower than  in a  binomial distribution.
If independents constitute the majority of the  voters,
the voting rate converges   at the same rate  as  that  in a  binomial distribution. 
If the independents vote for the correct candidate rather  than for the wrong  candidate,  the model does not include the case
  wherein the majority of the voters choose  the  wrong candidate. 
The herders affect only   the speed of the convergence;  they   do not affect the  voting rates  for  the  correct candidate.

Next,  we consider herders who always choose the candidate with   a  majority of the previous  votes, which is visible to them \cite{Hisakado3}.  We refer  to these herders as  digital herders.
Digital herders exhibit  stronger herd behavior  than  analog herders.
We obtained  exact solutions when the  voters  comprised a mix of digital herders and independents.
As the fraction of herders increases, the model features a phase transition beyond which a state where most voters make the correct choice coexists with one where most of them are wrong.
This phase transition  is referred to as information cascade transition.


Here,  we  discuss  a  voting model  
with two candidates, $C_0$ and $C_1$.
We set two types of voters--independents and herders.
The voting of independents  is  based on  their fundamental values.
They collect information independently.
On the other hand, 
the voting of herders  is  based on the number of  previous votes, which is visible to them.
In this study, we consider the case wherein  a voter can see all previous votes.

From experiments, we  observed that 
human beings exhibit a behavior between that of  digital and analog herders.
We obtained  the probability that a herder makes  a choice  under the influence of his/her prior voters' votes.
The probability  can be fitted by  a  $\tanh$  function.
If the difference between numbers  of voters for  candidates $C_0$ and $C_1$ is small, the probability that a herder chooses  the candidate receiving a majority  of the previous votes increases rapidly.
 If the difference between  numbers of voters for  candidates  $C_0$ and $C_1$  is large, the probability  becomes constant.
In this paper,  we discuss rich phases of the models
in  which herders   exhibit  behavior that can be fitted  to a   $\tanh$   function.
We  identify  two types of phase transitions--information cascade transition  and transition between super and normal diffusions.
Furthermore, we discuss the  phases of models that 
we  obtained from experiments \cite{Mori3}.



The remainder  of this paper is organized  as follows.
In section 2, we introduce our  voting model and 
  mathematically 
 define the two types of  voters--independents and herders. 
In section 3,
we derive a stochastic differential equation.
In section 4, we discuss 
information cascade transition  by using the  stochastic differential equation.
In section 5, we discuss the 
phase transition between normal and super diffusions   and we   demonstrate
the coexistence of these phase transitions.
In section 6, 
we verify  these transitions through numerical simulations.
In section 7, we   discuss  social experiments from the viewpoint of our models.
Finally, the conclusions are presented in section 8.

\section{Model}


We model the voting of two candidates, $C_0$ and $C_1$;
at  time $t$, $C_0$ and $C_1$ have  $c_0(t)$ and $c_1(t)$ votes, respectively.
In  each time step, one  voter  votes   for one  candidate;
 the voting is sequential.
Hence,  at time $t$,  the $t$th voter  votes,  after which the  total number of votes  is  $t$.
Voters are allowed to see   all the  previous votes for each candidate; thus, they are aware of   public perception.\footnote{The $t$th voter can see $c_0(t-1)$ and $c_1(t-1)$  votes at time $t$.}


There are  two types of voters--independents and herders;
 we assume   an  infinite number of voters.
The independents vote for $C_0$ and $C_1$
with  probabilities $1-q$ and $q$, respectively.
Their votes are independent  of  others' votes, i.e. 
their votes are based  on  their fundamental values.

Here, we set $C_0$ as the wrong  candidate and $C_1$ as the correct candidate in order  to validate the performance of the  herders.
We can  set $q\geq 0.5$  because we  believe that    independents vote for the correct  candidate $C_1$ rather   than for the wrong candidate $C_0$.
In other words, we assume that  the  intelligence of the independents is virtually   accurate.

\begin{figure}[h]
\includegraphics[width=120mm]{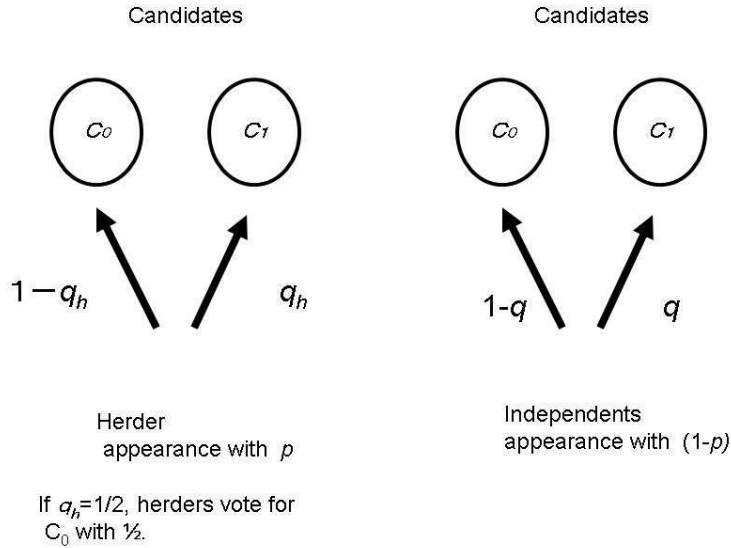}
\caption{Representation of our  model. Independents vote for $C_0$ and $C_1$
with  probabilities $1-q$ and $q$, respectively.
Herders vote for $C_0$ and $C_1$
with  probabilities $1-q_h$ and $q_h$, respectively.
We set the ratio of independents to 
herders as $(1-p)/p$.
 In the limit $\lambda\rightarrow \infty$,  a herder is  a digital herder. In the limit $\lambda=2$, a herder is similar to  an  analog herder. In general, the herder is a hybrid  between   an analog  herder and a digital herder with the parameter $\lambda$.  }
\label{model}
\end{figure}

On the other hand, the  herders' votes  are    based on the number of previous votes.
We use the following functions.
If the numbers of votes 
are $c_0(t)$ and $c_1(t)$  at time  $t$,
a  herder  votes for $C_1$ at time $t+1$  with the following  probability:
\begin{equation}
 q_h=\frac{1}{2} 
[ \tanh \lambda \{\frac{c_1(t)}{(c_0(t)+c_1(t))}-\frac{1}{2} \}+1 ].
\label{tanh}
\end{equation}
If  $C_1$ receives the  majority votes, i.e. $c_1(t)/(c_0(t)+c_1(t))>1/2$, the  ratio of votes  for $C_1$, denoted by  $q_h$, increases to $1$ exponentially. $\lambda$ reflects how the previous 
answers affect  a voter's choice.
If $\lambda$  is positive and large, the voter has high confidence in  the previous votes.
We use  function (\ref{tanh}) for our social experiments, and
it fits well to human behaviors in the experiments \cite{Mori3}.
In our experiments, we can estimate $\lambda=3.80$.
If  $c_0(t)=c_1(t)$, herders vote for $C_0$ and $C_1$
with the same  probability, i.e. $1/2$.
This model can  also be derived from  Bayes' theorem (see Appendix A).
In this case, $\lambda$ is not constant and  it increases as the number of votes increases, according to  the central limit theorem.
The case  is on the assumption that all voters are independents.
Hereafter, we  treat $\lambda$ as a parameter.(see figure \ref{model})

In the upper limit of $\lambda$, 
herders vote for  the   candidate with the majority votes.
If $c_0(t)>c_1(t)$, the  herders vote for  candidate $C_0$, whereas
if $c_0(t)<c_1(t)$,  they  vote for  candidate $C_1$.
These herders  are known as digital herders \cite{Hisakado3}.
We expand (\ref{tanh}) as follows:
\begin{equation}
 q_h\sim \frac{1}{2}+\frac{\lambda}{2}(\frac{c_1(t)}{(c_0(t)+c_1(t))}-\frac{1}{2})
-\frac{\lambda^3}{6}(\frac{c_1(t)}{(c_0(t)+c_1(t))}-\frac{1}{2})^3+\dots.
\label{ana}
\end{equation}
If $c_1(t)/(c_0(t)+c_1(t))\sim 1/2$ and $\lambda=2$,  a herder votes  for $C_1$   with  a  probability of 
$ c_1(t)/(c_0(t)+c_1(t))$.
These  herders are known as analog herders with $q=1/2$ \cite{Hisakado2}.
Therefore,  the herders who vote with   probability  (\ref{tanh}) are 
a hybrid  of  analog and digital herders.

The independents and herders appear randomly and vote.
We set the ratio of independents to 
herders as $(1-p)/p$.
In this study, we mainly focus on  the upper  limit of $t$.
This refers to  the voting of infinite voters.

The evolution  equation for  candidate $C_1$ is
\begin{eqnarray}
P(k,t)&=&pq_hP(k-1,t-1)
+p(1-q_h)P(k,t-1)
\nonumber \\
& &
+(1-p)qP(k-1,t-1)
+(1-p)(1-q)P(k,t-1).
\label{m1}
\end{eqnarray}
Here, $P(k,t)$ is
the distribution of 
the number of votes $k$  at time $t$ for  candidate $C_1$.
The first and  second terms of (\ref{m1}) denote  the votes of  the
herders;
the third and  fourth terms denote the votes of  the
independents.

\section{Stochastic Differential Equation}

To investigate   long-ranged correlations, we 
analyze in the limit $t \rightarrow \infty$.  
We can rewrite  (\ref{m1}) as
\begin{eqnarray}
c_1(t)=k \rightarrow k+1:
 P_{k,t}&=&\frac{p}{2} 
[ \tanh \lambda (\frac{k}{t-1}-\frac{1}{2} )+1 ]+(1-p)q.
\label{pd}
\end{eqnarray}
In the scaling limit $t=c_0(t)+c_1(t)\rightarrow \infty$,
 we define
\begin{equation}
\frac{c_1(t)}{t}
\Longrightarrow Z.
\label{Z}
\end{equation}
$Z$ is the ratio of voters who vote for $C_1$.


We define  a new variable $\Delta_t$ such that
\begin{equation}
\Delta_t=2c_1(t)-t=c_1(t)-c_0(t).
\label{d}
\end{equation}
We change the notation from $k$ to $\Delta_t$ for convenience.
Then, we have $|\Delta_t|=|2k-t|<t$.
Thus, $\Delta_t$ holds  within   $\{-t,t\}$. 
Given $\Delta_t=u$, we obtain a random walk model:
\begin{eqnarray}
\Delta_t&=&u \rightarrow u+1  :P_{\frac{u+t}{2},t}=
\frac{p}{2} 
[ \tanh  \{\frac{\lambda u}{2(t-1)}\}+1 ]
+(1-p)q,
\nonumber \\
\Delta_t&=&u \rightarrow u-1  :Q_{\frac{u+t}{2},t}=1-P_{\frac{u+t}{2},t}.
\nonumber
\end{eqnarray}
We now  consider the continuous limit $\epsilon \rightarrow 0$,
\begin{eqnarray}
X_\tau&=&\epsilon\Delta_{[t/\epsilon]},
\nonumber \\
P(x,\tau)&=&\epsilon P(\Delta_t/\epsilon,t/\epsilon),
\end{eqnarray}
where $\tau=t/\epsilon$ and $x=\Delta_t/\epsilon$.
Approaching the continuous limit, we can obtain the Fokker-Planck equation (see Appendix B):
\begin{equation}
\textrm{d}X_\tau=[(1-p)(2q-1)+p 
\tanh  (\frac{\lambda X_\tau}{2\tau})]\textrm{d}\tau+\sqrt{\epsilon}.
\label{ito0}
\end{equation}
Here, we change  the variable  $X_\tau$ to $Y_\tau$ by using the expression,
\begin{equation}
pY_\tau=X_\tau-(1-p)(2q-1)\tau.
\label{y}
\end{equation}
We rewrite (\ref{ito0})  by using $Y_\tau$:
\begin{equation}
\textrm{d}Y_\tau=\tanh  \frac{p\lambda}{2\tau}(Y_\tau+\frac{(2q-1)(1-p)}{p}\tau)\textrm{d}\tau+\sqrt{\epsilon}.
\label{ito2}
\end{equation}
Using (\ref{Z}), (\ref{d}), and (\ref{y}),
we can obtain the relations of the  variables 
\begin{equation}
2Z-1=\frac{X_{\infty}}{\tau}=\frac{pY_{\infty}}{\tau}+(1-p)(2q-1).
\label{re}
\end{equation}

\section{Information Cascade Transition}

In this section, we discuss the information cascade transition.
We observed  this transition in the case of digital herders \cite{Hisakado3}.
We are interested in the behavior in the   limit $\tau\rightarrow \infty$.
We consider the solution $Y_\infty\sim\tau^{\alpha}$, where $\alpha\leq1$,  
since  $\tanh x\leq1$.
The slow   solution is  $Y_\infty\sim\tau^{\alpha}$, where $\alpha<1$ is hidden by
the fast    solution $\alpha=1$ in the upper  limit of $\tau$.
Hence, we  can assume a stationary solution as
\begin{equation}
Y_\infty=\bar{v}\tau,
\label{h}
\end{equation}
where $\bar{v}$ is constant.
Substituting  (\ref{h})  into (\ref{ito2}), we can obtain
\begin{equation}
\Delta Y_\infty=\tanh \frac{p\lambda}{2}(\bar{v}+\frac{(2q-1)(1-p)}{p}) \Delta \tau=\bar{v}\Delta \tau.
\end{equation}
The second equality is obtained  from (\ref{h}).
Then, we obtain  the equation
\begin{equation}
\bar{v}=\tanh \frac{p\lambda}{2}(\bar{v}+\frac{(2q-1)(1-p)}{p}).
\label{ising}
\end{equation}
This is the  equation of state for Ising model.
The second term on  the  RHS of (\ref{ising}) corresponds to the 
external field.
In the cases  of $q=1/2$ and $p=1$, 
the external field disappears.
If $\lambda>2$,  a phase transition occurs  in the range $0\leq p\leq1$.
As the number  of herders increases, the model features   a phase transition beyond which  a state where most voters make the correct  choice coexists with one where most of them are wrong.
We refer to this transition as information cascade transition \cite{Hisakado3}.

\begin{figure}[h]
\includegraphics[width=120mm]{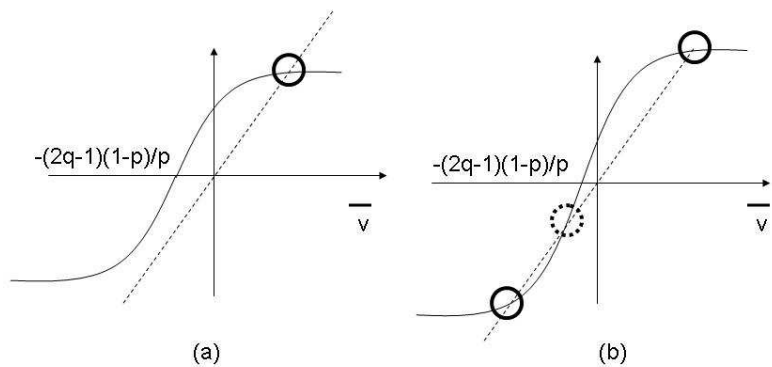}
\caption{Solutions of self-consistent equation (\ref{ising}). It is  the state equation of Ising model. (a) $p\leq p_c$  and (b) $p>p_c$. Below the critical point $p_c$, we can obtain one solution (a). We refer to  this phase  as the  one peak phase. In contrast, above the critical point, we obtain three solutions. Two of them are stable and one is unstable (b). We refer to  this phase as the  two-peaks phase. In Ising model, the physical solution is  the right side one only. }
\label{Ising}
\end{figure}

Equation (\ref{ising}) admits one solution  below the critical point $p\leq p_c$  (see figure  \ref{Ising}(a))
 and  three  solutions for $p>p_c$  (see figure \ref{Ising}(b)).
When  $p\leq p_c$, we refer to the phase as the  one peak phase.
When  $p>p_c$, the  upper  and lower solutions 
are  stable;
on the other hand, 
the intermediate solution is   unstable.
Then, the  two  stable solutions attain a  good and bad equilibrium,  
and  the distribution becomes the sum of the two Dirac measures.
We refer to this phase as  two-peaks phase.
In  Ising model with external fields, the good  equilibrium is only  a physical solution.
This  is the difference between our voting model and Ising model.
Here, we discuss  two particular cases.

\noindent{\bf (1) Digital herder case: $\lambda=\infty$ }

In this case,  the herders are considered to be   digital herders.
As shown in  figure  \ref{Ising}, the $\tanh$ function rises  vertically at $\bar{v}=-\frac{(2q-1)(1-p)}{p}$.
A phase transition occurs at $p_c=1-\frac{1}{2q}$.
When $p\leq p_c$, we can obtain only one solution  $\bar{v}=1$.
Using (\ref{re}), we obtain  the ratio of voters for $C_1$ as 
$Z=p+(1-p)q$.
When $p> p_c$, we can obtain  two stable  solutions: $\bar{v}=1$ and
$\bar{v}=-1$.
The solution $\bar{v}=-\frac{(2q-1)(1-p)}{p}$ is unstable.
Using (\ref{re}), the ratio of voters for $C_1$ is obtained as
$Z=p+(1-p)q$ and $Z=q(1-p)$.
$Z=p+(1-p)q$ shows   a good equilibrium, whereas 
 $Z=q(1-p)$ shows  a bad equilibrium.
This  conclusion is consistent with the exact solutions \cite{Hisakado3}.

\noindent{\bf (2) Symmetric independent voter case: $q=1/2$ }

In this case, the external fields are  absent.
The self-consistent equation (\ref{ising}) becomes
\begin{equation}
\bar{v}=\tanh \frac{p\lambda}{2}(\bar{v}).
\label{ising2}
\end{equation}
As shown   in figure  \ref{Ising}, the $\tanh$ function rises  at $\bar{v}=0$.
If $\lambda\leq 2$, there is only one solution $\bar{v}=0$  in all regions of  $p$.
In this case, $Z$ has only one peak, at $0.5$, 
which indicates
 the one peak phase.
 When herders are  analog herders, we do not observe information cascade transition \cite{Hisakado3}.
We  observe only  super and normal transitions (see section 5).
On the other hand, if $\lambda>2$,
there are  two stable solutions and  an unstable solution $\bar{v}=0$ above $p_c$.
The   votes ratio for $C_1$ attains 
a good  or   bad equilibrium.
This is the so-called spontaneous symmetry breaking.
In one sequence, $Z$ 
 is  taken as $\bar{v}p/2+1/2$ in the case of  a good equilibrium,  or
as  $-\bar{v}p/2+1/2$ in the case of   a bad equilibrium,
where $\bar{v}$ is the solution of (\ref{ising2}).
This indicates  the two-peaks phase, and 
the critical point  is $p_c=2/\lambda$.


\section{Phase Transition of Super  and Normal Diffusion Phases} 

In this section, we consider the phase transition of convergence.
This type of transition has been  studied when herders are analog \cite{Hisakado2}.
The analog herders exhibit    weaker  herd behavior  than digital herders.
Depending on the convergence behavior, there are three phases.
We expand $Y_\tau$ around the solution $\bar{v}$.
\begin{equation}
Y_\tau=\bar{v}\tau+W_\tau.
\label{w}
\end{equation}
Here, we set $Y_\tau\gg W_\tau$.
This indicates  $\tau\gg 1$.
We rewrite (\ref{ito2}) using (\ref{w}) as follows:
\begin{eqnarray}
\textrm{d}Y_\tau&=&
\bar{v}\textrm{d}\tau+\textrm{d}W_\tau
\nonumber \\
&=&\tanh [ \frac{p\lambda}{2\tau}(\bar{v}\tau+\frac{(2q-1)(1-p)}{p}\tau)+\frac{p\lambda}{2\tau}W_\tau]\textrm{d}\tau+\sqrt{\epsilon}
\nonumber \\
&\sim&
\bar{v}\textrm{d}\tau+\frac{p\lambda}{2\tau\cosh^2 \frac{p\lambda}{2}(\bar{v}+\frac{(2q-1)(1-p)}{p}) }W_\tau\textrm{d}\tau+\sqrt{\epsilon}
\nonumber \\
&=&
\bar{v}\textrm{d}\tau+\frac{p\lambda(1-\bar{v}^2)}{2\tau}W_\tau\textrm{d}\tau+\sqrt{\epsilon}
.
\label{ito3}
\end{eqnarray}
We use  relation (\ref{ising}) and  consider  the first term of the expansion.
Hence, we can obtain
\begin{equation}
\textrm{d}W_\tau=\frac{p\lambda(1-\bar{v}^2)}{2\tau }W_\tau\textrm{d}\tau+\sqrt{\epsilon}.
\label{ito4}
\end{equation}

From Appendix C,
we can obtain   the phase transition of convergence.
The critical point $p_{vc}$ is 
the solution of 
\begin{equation}
p_{vc}=\frac{1}{\lambda(1-\bar{v}^2)},
\label{vc} 
\end{equation}
and (\ref{ising}).


As shown in   figure  \ref{Ising},   at the critical point,  the gradient of the  tangent line at $\bar{v}$ is  $1/2$.
If the gradient of the  tangent line at $\bar{v}$ is  under  $1/2$,
the distribution converges as   in a binomial distribution.
We define $\gamma$ as $Var(Z)=\tau ^{-\gamma}$, where $Var(Z)$ is the  variance of $Z$.
The voting rate for $C_1$ converges as $Var(Z)=\tau ^{-1}$: $\gamma=1$. 
We refer to  this phase as a normal diffusion  phase.
If the gradient of the tangent line at $\bar{v}$ is  $1/2$ or above $1/2$,
    the voting rate   converges at a speed slower  than that in  a binomial distribution.
We refer to  these phases as  super diffusion  phases.
In one phase,  the voting rate for $C_1$ converges to  $\log (\tau)/\tau$,
and in 
the other,   the voting rate converges to $\gamma=p/p_{vc}-2$.

\noindent{\bf (1) Digital herder case: $\lambda=\infty$ }

In this case, the herders are considered to be   digital herders.
In  the one peak phase, where  $p\leq p_c$, the only solution is  $\bar{v}=1$.
Equation (\ref{ito4}) represents the Brownian motion.
The gradient of the  tangent line at $\bar{v}$ is  $0$.
Hence, the distribution converges as   in a binomial distribution.
In the  two-peaks phase, where $p> p_c$,
the solutions are   $\bar{v}=\pm 1$.
The gradient of  the tangent line at $\bar{v}$ is  $0$.
In each case, the distribution converges  as in    a binomial distribution.
Hence, in all regions, 
the distribution converges as  in   a binomial distribution.
In this limit, the phase transition   of the convergence disappears and  only  information cascade transition is observed.  

\noindent{\bf (2) Symmetric independent voter case: $q=1/2$}

We  consider  the   case $\lambda>2$.
In this case, we observe   information cascade transition.
If $\lambda\leq 2$, we do not observe   information cascade transition and   we can only observe  a part of the phases, as described  below.

In  the one peak phase $p\leq p_c=2/\lambda$, the only solution is  $\bar{v}=0$.
$p_c$ is the critical point  of the  information cascade transition. 
The first critical point of convergence  is $p_{vc1}=1/\lambda$.
When  $p\leq p_c$, $p_{vc1}$ is the solution of (\ref{ising2}) and (\ref{vc}).
If  $0<p<p_{vc1}$, the voting rate for $C_1$ becomes $1/2$, and  the distribution converges as  in a binomial distribution.
If $p_{c}>p\geq p_{vc1}$,
   candidate $C_1$  gathers $1/2$ of all the  votes in the scaled distributions, too. 
However, the voting rate   converges slower  than  in  a binomial distribution.
We refer to these phases as  super   diffusion phases.
There are two phases, $p=p_{vc1}$ and $p_{c}>p>p_{vc1}$;
these  phases differ  in terms of their convergence speed.

Above $p_c$, in  the two-peaks phase,  we can obtain two stable solutions that  are not $0$.
At $p_c$, $\bar{v}$ moves from $0$ to one of these two stable solutions.
In one voting sequence, the votes converge to one of these stable  solutions.
If  $p_{c}<p\leq p_{vc2}$, the voting rate for $C_1$ becomes $\bar{v}p/2+1/2$ or $-\bar{v}p/2+1/2$, and  the convergence occurs  at a rate slower
than that   in a binomial distribution. 
Here, $\bar{v}$ is the solution of (\ref{ising2}).
We refer to  this phase as a  super diffusion phase.
 $p_{vc2}$ is the second critical point of convergence from  the super to the normal diffusion phase,  and it is the solution of the simultaneous equations 
(\ref{ising2}) and (\ref{vc}) when  $p> p_c$.
We can estimate $p_{vc2}\sim\frac{2.5}{\lambda}$ by approximation.\footnote{
Using $\tanh x\simeq x-1/3x^3$, 
 equation (\ref{ising2}), and the condition of  critical point,
 we can obtain  $p_{vc2}\simeq 2.5/\lambda$.}
In fact, with higher terms, we can estimate $p_{vc2}\sim\frac{2.7}{\lambda}$ (see figure \ref{pd}(a)).
On the other hand, if  $p>p_{vc2}$, the voting rate for $C_1$ becomes $\bar{v}p/2+1/2$ or $-\bar{v}p/2+1/2$, too.
But the distribution converges as   in a binomial distribution. This  is a  normal diffusion phase.
A total of  six  phases can be observed.

\begin{figure}[h] 
\begin{center} 
\begin{tabular}{c} 
\begin{minipage}{0.5\hsize}
\begin{center} 
\includegraphics[clip, width=6.5cm]{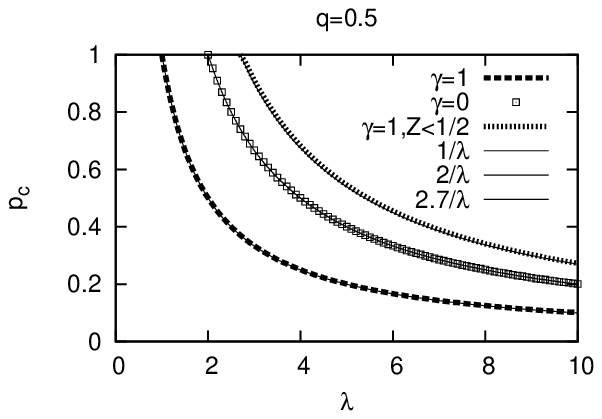}
\hspace{1.6cm} (a)
\end{center} 
\end{minipage} 
\begin{minipage}{0.5\hsize} 
\begin{center} 
\includegraphics[clip, width=6.5cm]{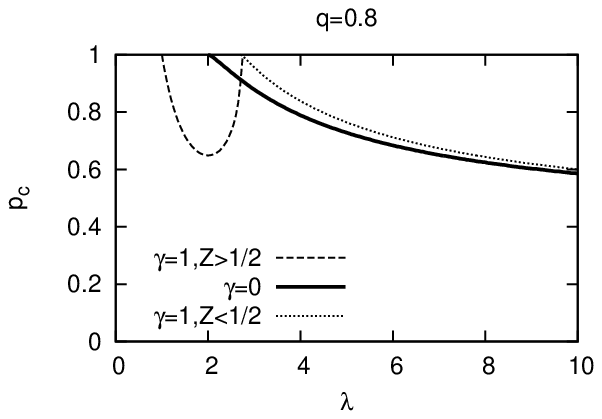} 
\hspace{1.6cm} (b) 
\end{center} 
\end{minipage} 
 \end{tabular} 
\caption{Phase diagram in  space $\lambda$ and $p$ when  (a) $q=0.5$ and (b) $q=0.8$.
$\gamma$  denotes the speed of convergence.
$\gamma=0$ is the boundary between  the one peak phase and the  two-peaks phase.
The region below  $\gamma=0$ is the one peak phase. 
The region above $\gamma=0$ is  the two-peaks phase.
$\gamma=1$  is the between  the  normal  diffusion phase and  the super diffusion phase.
In the one peak phase, the  region below $\gamma =1$ is the normal diffusion phase.
 In  the two-peaks phase, two solutions are obtained.
Each solution has  a different speed of convergence, generally.
When $q=0.5$, there is no difference  between the two solutions and  the  region below $\gamma =1$ is the super  diffusion phase.
When $q=0.8$, there is  a difference  between the  good   and the  bad equilibriums.
$Z>1/2$ is the good equilibrium and $Z<1/2$ is the bad equilibrium.
The left side of $\gamma=1$, $Z>1/2$ in the  two-peaks phase   is the phase in which
a good  equilibrium is a   super diffusion phase.
The left side of  $\gamma=1$, $Z<1/2$  is the  phase
in which   a bad equilibrium  is a super diffusion phase.} 
\label{pd} 
\end{center} 
\end{figure}

As discussed above, when $q=0.5$, 
there is no difference between  good  and bad equilibriums.
However, when   $q=0.8$, we observe a  difference.
$Z>1/2$ indicates the good equilibrium and $Z<1/2$ indicates the bad equilibrium.
Figure \ref{pd} shows
the phase diagram in  space $\lambda$ and $p$ when (a) $q=0.5$ and (b) $q=0.8$.
$\gamma$ is the speed of convergence.
$\gamma=0$ is the boundary between the one peak phase and  the two-peaks phase.
The region below  $\gamma=0$ is  the one peak phase and that
above $\gamma=0$ is the  two-peaks phase. 
$\gamma=1$  is  between the  normal  diffusion phase and the super diffusion phase.
In the one peak phase, the  region below $\gamma =1$ is the normal diffusion phase
and  that above $\gamma =1$ is the super diffusion phase.
 In the  two-peaks phase, two solutions can be observed.
When $q=0.5$, we do not observe a  difference  between the two solutions
 and  the  region above $\gamma =1$ is the normal diffusion phase.
When $q=0.8$, we observe a  difference  in the speed of convergence between   good  and bad equilibriums.
$Z>1/2$ is the good equilibrium and $Z<1/2$ is the bad equilibrium.
The left side of $\gamma=1$, $Z>1/2$ in the  two-peaks phase   is the phase in which
a good  equilibrium is   the super diffusion phase.
The left side of  $\gamma=1$,  $Z<1/2$  is the  phase
in which   a bad equilibrium is the  super diffusion phase.

When  $q=0.5$, the curves $\gamma=1$, $Z>1/2$ in  the one peak phase and 
the two-peaks phase are separated by the curve $\gamma=0$.
When  $q\neq0.5$, the curves $\gamma=1$, $Z>1/2$ in the  one peak phase and the two-peaks phase are deformed and meet on the curve $\gamma=0$. 
Hence, the region surrounded by 
$\gamma=1$, $Z>1/2$ is the super diffusion phase of the  good equilibrium.

Finally,  we comment  on the  analog herder case.
As discussed in section 2, if we set $\lambda=2$ and use only the linear terms in (\ref{tanh}), we can obtain the  analog herder case. 
In this limit, we can solve (\ref{ising}) and obtain  $\bar{v}=2q-1$.
Equation (\ref{ising}) has only one solution in entire region $p$.
Hence, we  do not observe information cascade transition, which is consistent with previous conclusions \cite{Hisakado2}.
On the other hand, in  super and normal diffusion phase transitions, 
$p_{vc1}=1/2$  does not depend on $q$,  because the gradient of the  linear term  is constant.

\section{Numerical Simulations}

To confirm the analytical results, we performed  numerical integration
of the master equation (\ref{m1}).
We  perform  simulations  for the  symmetric  independent voter case, i.e. $q=1/2$.

%





\begin{figure}[h] 
\begin{center} 
\begin{tabular}{c} 
\begin{minipage}{0.5\hsize}
\begin{center} 
\includegraphics[clip, width=6.5cm]{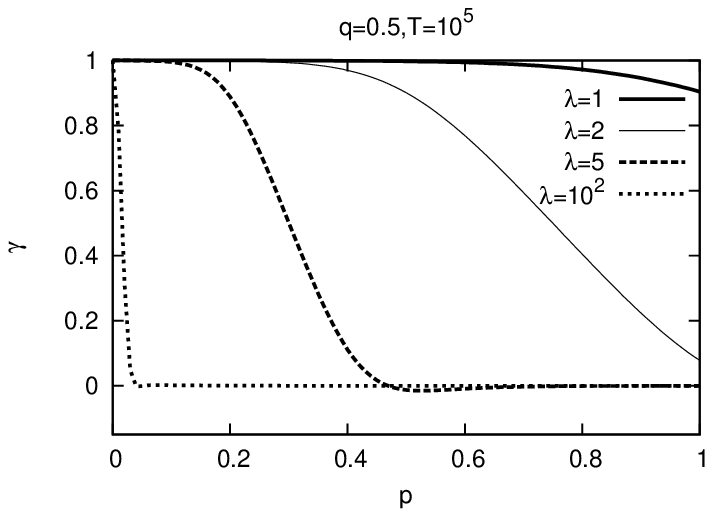}
\hspace{1.6cm} (a)
\end{center} 
\end{minipage} 
\begin{minipage}{0.5\hsize} 
\begin{center} 
\includegraphics[clip, width=6.5cm]{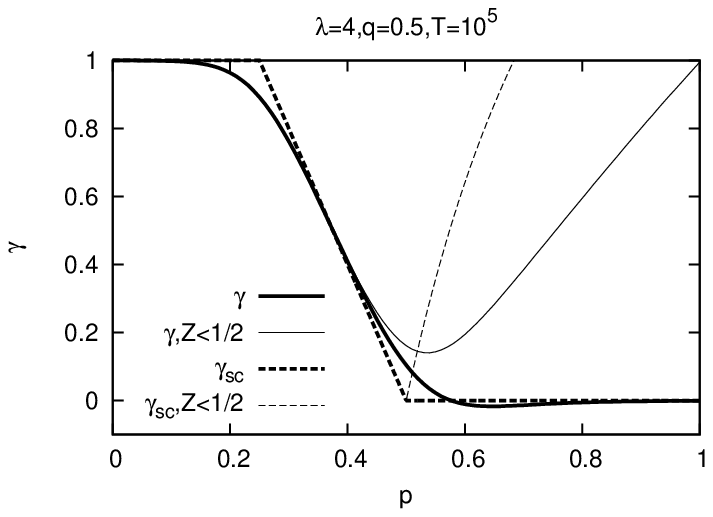} 
\hspace{1.6cm} (b) 
\end{center} 
\end{minipage} 
 \end{tabular} 
\caption{Convergence of  distribution in the one peak phase and the two-peaks phase for the $q=0.5$ case. The horizontal axis  represents the ratio of herders $p$, and the vertical axis represents the speed of convergence.
$\gamma=1$ is the normal  phase,  and $0<\gamma<1$ is the super diffusion phase.
(a)  is for the one peak  phase.
The critical point of the convergence of transition  from normal diffusion to super diffusion is $p_{vc1}\sim 1/\lambda$, and the critical point of information cascade  transition is $p_{c}\sim 2/\lambda.$ 
 (b)  is for the two-peaks phase. The solid lines are obtained by numerical simulations.
$\gamma,Z<0$ is  the speed of convergence of the region $Z<0$, 
which indicates a  bad equilibrium.
The dotted line $\gamma _{sc}$    represents  the  theoretically obtained curve and 
$\gamma _{sc}, Z<0$  represents  the  theoretically obtained curve  for a bad equilibrium.
The critical point of information cascade  transition is $p_{c}\sim 0.5$ and
the critical point of the convergence of transition is  $p_{vc1}\sim 0.25$ and
$p_{vc2}\sim 0.675$. } 
\label{syml} 
\end{center} 
\end{figure}

\noindent{\bf (1) One peak phase: $q=1/2,\lambda=4$}

Figure \ref{syml}(a) shows the convergence of the distribution in the one peak phase.
We integrated the master equation up to $t=10^5$.
As discussed in previous sections, 
the critical point of the convergence of  this transition is $p_{vc1}\sim 1/\lambda$ and the critical point of information cascade transition is $p_{c}\sim 2/\lambda$.
At the critical point of information cascade transition, the distribution splits into two  and the exponent $\gamma$  becomes $0$.\footnote{
Here, we estimate $\gamma$ from the slope of ${\rm Var}(Z(t))$ as
$
\gamma=\log \{ {\rm Var}(Z(t-\Delta t)) /{\rm Var}(Z(t))\} /
\log \{t/(t-\Delta t)\}.
$
}



\noindent{\bf (2) Two-peaks phase: $q=1/2,\lambda=4$}

Figure \ref{syml}(b) shows the convergence of the distribution in the two-peaks phase.
We consider the case wherein $q=0.5$ and  $\lambda=4$.
In this phase,  a sequence of voting converges to one of the peaks.
We can not  determine  which peak is  chosen in finite time.
We divide  the distribution into  two parts,  $Z>0$ or $Z<0$, which correspond to two regions around the peaks.

Above the  critical point of information cascade transition  $p_{c}\sim 2/\lambda=0.5$,
 the speed of the convergence in the region $Z<0$  is slower than that  in a  binomial distribution.\footnote{In the case $q=0.5$,   the convergence in the region $Z>0$ is  identical to that in $Z<0$ because of  the symmetry. We  consider only  $Z<0$ here.} 
This indicates the super diffusion phase.
In the super diffusion phase, the distance between the two peaks is short and  the influence of the  other peak reduces the convergence speed.
After the transition of convergence to $p_{vc2}=0.675$, the speed of the convergence is as in a  binomial distribution $\gamma=1$.
This  indicates  the normal diffusion phase.

The critical point  of convergence, $p_{vc2}$, in  the two-peaks phase is larger  than that  predicted  by    theory. 
We calculated the variance  in the region $Z<0$  using the data from $t=9.99\times10^4$ to $t=10^5$.
The region may be too wide for  use in   our assumptions,  which we described   in the previous section (\ref{w}).
To fill  the gap between the numerical calculations and  theory, 
 we need to perform simulations for a large $t$.

\section{Social Experiments}
We conducted simple social experiments for our model.
In 2010, we framed  100 questions, each  with two choices--knowledge and  no knowledge.
31 participants answered  these questions sequentially
 in one group.
We performed experiments  with two groups.
In 2011, we framed  120 questions and 
52 participants answered  these questions sequentially.
We  again performed  experiments with two groups.

First, they answered the questions without having  any information about the others'  answers,
 i.e.  their  answers were   based on  their own knowledge.
Those  who   knew the answers    selected the correct answers.
Hence, we can set $q=1$.
Those who    did not  know the answers   selected the correct answers  with  a probability of $0.5$. 
Next,  the  participants were allowed to see   all   previous participants' answers.
Those who  did  not  know the answers  referred  to this  information.
We are interested in knowing  whether they referred to this  information as digital  or  analog herders.

From the experiments, we can obtain macroscopic  behavior $q_h$ \cite{Mori3}.
The voters can see all  the previous votes, and 
we  could fit the plot by the following functional form:
\begin{equation}
q_h=\frac{1}{2} 
[ a \tanh \lambda \{\frac{c_1(t)}{(c_0(t)+c_1(t))}-\frac{1}{2} \}+1 ].
\label{tanh2}
\end{equation} 
The difference between (\ref{tanh}) and (\ref{tanh2}) is the constant $a$.
The parameter $a$ denotes the net  ratio  of  the herder that reacts positively to the previous votes.
We  estimated the parameters $\lambda=3.80$ and $a=0.761$ form the experiments.

We can map  (\ref{tanh2}) to (\ref{tanh})  as follows:
\begin{eqnarray}
P(k,t)&=&pq_h+(1-p)=
\frac{1}{2}p 
[ a \tanh \lambda \{\frac{c_1(t)}{(c_0(t)+c_1(t))}-\frac{1}{2} \}+1 ]+(1-p)
\nonumber \\
&=&
\frac{1}{2}\tilde{p} 
[  \tanh \lambda \{\frac{c_1(t)}{(c_0(t)+c_1(t))}-\frac{1}{2} \}+1 ]+(1-\tilde{p})\tilde{q}, 
\end{eqnarray}
where
$\tilde{p}=pa$ and $\tilde{q}=1-1/2\cdot p(1-a)/(1-pa)$.
Using this mapping and conclusions from previous sections, we can theoretically  estimate the conclusions for a large limit of $t$.

\begin{figure}[h] 
\begin{center} 
\begin{tabular}{c} 
\begin{minipage}{0.5\hsize}
\begin{center} 
\includegraphics[clip, width=6.5cm]{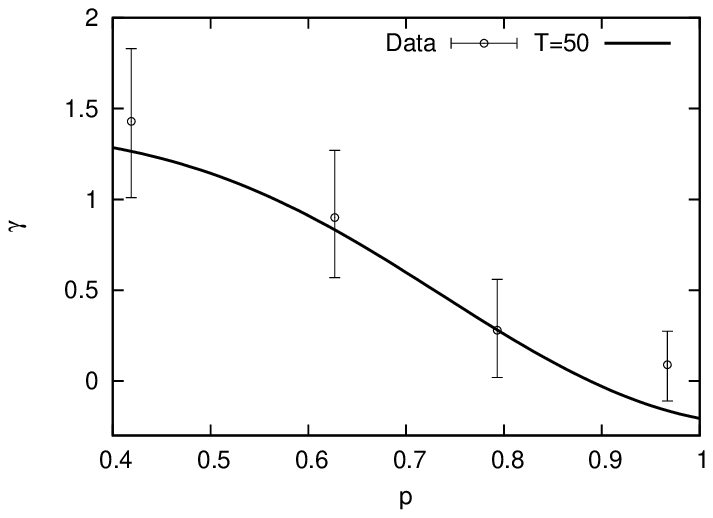}
\hspace{1.6cm} (a)
\end{center} 
\end{minipage} 
\begin{minipage}{0.5\hsize} 
\begin{center} 
\includegraphics[clip, width=6.5cm]{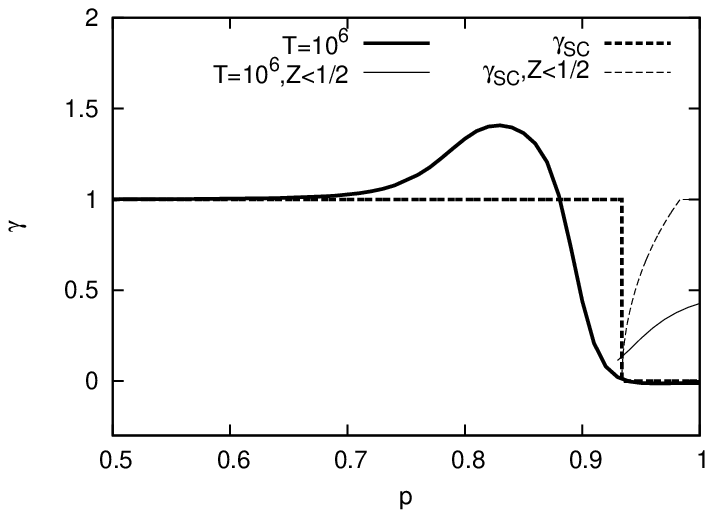} 
\hspace{1.6cm} (b) 
\end{center} 
\end{minipage} 
 \end{tabular} 
\caption{Behavior of convergence as given by the plot of
$p$ vs. $\gamma$. 
The horizontal axis  represents the ratio of herders, $p$, and the vertical axis represents the speed of convergence.
$\gamma=1$ is the normal phase,   and $0<\gamma<1$ is the super diffusion phase. 
(a) is  for the experimental data  and simulation results at $t=50$.
(b) is for simulation results at $t=10^6$ and theoretical results in the limit $t \rightarrow \infty$.  
On the basis of  the theoretical results, we show the convergence of $Z$ as $\gamma_{sc}$ and that 
of $Z<1/2$ as  $\gamma_{sc}, Z<1/2$.
$\gamma_{sc}, Z<1/2$ shows the theoretical convergence for a  bad  equilibrium.
$T=10^6$ shows the simulation  curve of  the convergence of $Z$.
$T=10^6,Z<0$ shows the simulation  curve for a bad equilibrium.
Above the threshold $p_c=0.934$,  we observe in the two-peaks phase.} 
\label{exp} 
\end{center} 
\end{figure}

Figure \ref{exp}(a)
 shows the experimental data and  simulation results.
$\gamma=1$ is the normal  and $0<\gamma<1$ is the super diffusion phase. 
In the experimental data, when  the number of voters is  $t=50$,
we can confirm  that $\gamma$ monotonically decreases.
For the simulations, we set $t=50$, and 
we perform  numerical integration
of the master equation (\ref{m1}), as described in the previous section.
The data points observed in the experiments are   on the simulation curve without $p\sim 1$.

Figure \ref{exp}(b)
 shows the simulation 
 and 
theoretical  results to study the asymptotic behavior of convergence.
For $T=10^6$, we obtain   the simulation  curve of the convergence of $Z$.
In  $T=10^6$, represents a simulation curve,  $\gamma$  decreases from $1$ to
$0$ at $p_c\sim 0.9$.
This indicates the phase transition from the 
one peak phase to  the two-peaks phase. 

Using the conclusions of sections 4 and 5,
we can theoretically estimate the conclusions  for a large $t$ limit.
At $p_c=0.934$,
we observe  information  cascade transition.
$\gamma_{sc}$ is the theoretical curve that shows 
 information cascade transition at $p_c$.
We observe a   difference between the simulation curve $T=10^6$ and 
 the theoretical curve $\gamma_{sc}$.
This difference can  be reduced by performing a simulation for a large $t$.

In  the one peak phase, below $p_c$,
the peak  converges   as normal.
Above $p_c$,   we observe two peaks.
The good equilibrium converges  as normal.
On the other hand,  bad equilibrium converges  slower  than normal.
$\gamma_{sc},Z<0$,  represents a  theoretical curve, shows 
this phenomenon.
At $p_{cv2}=0.983$,
we observe the phase transition of normal and super diffusions, for the bad equilibrium.
Above $p_{cv2}$,
 both peaks converges as normal.
$T=10^6,Z<0$ is the simulation  curve  for the bad equilibrium.
We can observe the increase in  the  convergence speed  for  the  bad equilibrium.
The difference between the simulation  and the theoretical results  has been described   in the previous section; this difference  may  also  be reduced by performing a simulation for a large $t$.

\section{Concluding Remarks}

We investigated a voting model  that is similar to 
a Keynesian beauty contest.
In the continuous limit, we could obtain  stochastic differential equations.
The model has two kinds of  phase transitions.
One is information cascade transition,
which is similar to the  phase transition of Ising model.
In fact, we showed that the stationary condition of our model  is same as the equation of state for Ising model.
As the  herders increased, the model featured  a phase transition beyond which   a state where most voters make the  correct choice coexists with one where most of them are wrong.
In this transition, the distribution 
of votes changed from the  one peak phase to the  two-peaks phase.
These two peaks were the two  stable solutions out of  the three solutions.
In Ising model with  an external field, there is only one physical solution out  of the three   solutions.

The other transition  was  the transition of the convergence between super and normal diffusions.
In  the  one peak phase, if herders increased, the variance  converged slower than  in a binomial distribution.
This is the transition 
from normal diffusion to  super diffusion.
In  the two-peaks phase, the sequential voting  converged to one of the two peaks.
When $q=0.5$, 
this is   spontaneous symmetry breaking.
In the two-peaks phase, if the  herders increased,
the variance converged as in a binomial distribution.
This is the transition 
from super  diffusion to  normal diffusion, 
which  is opposite to the transition  in the  one peak phase.
In other words,
the super diffusion phase is sandwiched between  normal diffusion phases and
 is divided by information cascade transition.

We determined  the microspecific behavior
from social experiments.
 Using this experimental data, we confirmed the two kinds of phase transitions
by performing numerical simulations and  conducting analytical studies 
in the upper $t$ limit.

If   the ratio of herders  is smaller than $p_c$, we can
 distribute correct answers to herders that do not have information by  using this  voting system \cite{hayek}.
On the other hand,
 if the  ratio of herders is  larger than $p_c$, the system is similar to  a Keynesian beauty contest;   many voters are  herders and  there is a case in which  more than half the  voters make the wrong decision \cite{Keynes}.
In the case  $p<p_c$,  the advantage  of this system  is evident, and 
in the case  $p>p_c$, the  weakness is observed.
These are  two sides of the same coin and  are  
brought out  by information cascade  transition.

\begin{figure}[h] 
\begin{center} 
\begin{tabular}{c} 
\begin{minipage}{0.5\hsize}
\begin{center} 
\includegraphics[clip, width=6.5cm]{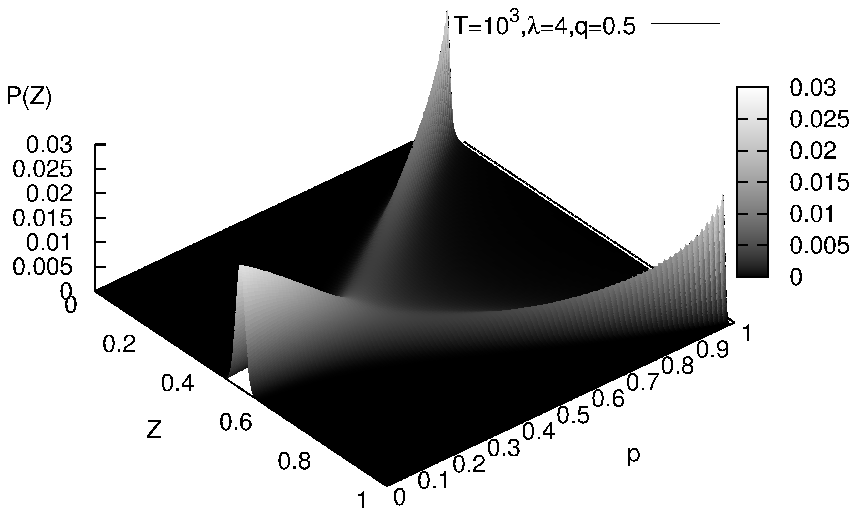}
\hspace{1.6cm} (a)
\end{center} 
\end{minipage} 
\begin{minipage}{0.5\hsize} 
\begin{center} 
\includegraphics[clip, width=6.5cm]{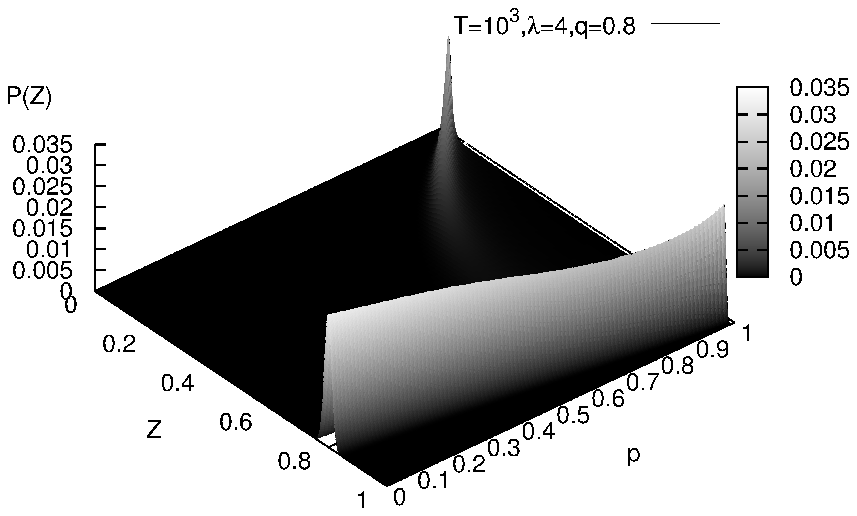} 
\hspace{1.6cm} (b) 
\end{center} 
\end{minipage} 
 \end{tabular} 
\caption{Distributions for  (a) $q=0.5$ and (b) $q=0.8$, 
 when  the number of  votes is $10^3$. 
The three axes are  $Z$, $p$, and the frequency $P(Z)$.
As $p$ increases, 
the distribution splits into  two  in both cases.
In the region near $p_c$, both the one peak and the  two-peaks phases 
exhibit   super diffusion.
In this phase, the speed of  convergence is slower than that   in the normal diffusion phase.
We can confirm  that the width of the distribution is greater  than that  in the region far from $p_c$.} 
\label{co} 
\end{center} 
\end{figure}

Figure \ref{co}
 shows  the  distributions when (a) $q=0.5$ and (b) $q=0.8$.
As $p$ increases, 
the distribution  splits into  two  in both cases.
We can  think of this as  splitting of  a particle    into two parts by the interactions.
In the region near $p_c$, both  the   one peak and the    two-peaks phases 
exhibit   super diffusion.
In this phase, the speed of  convergence is slower than that   in the normal diffusion phase.
We believe  that the particles are deformed by the interactions.
When  $q=0.5$, a particle is divided continuously.
When  $q=0.8$,  a second particle appears far from  the original particle and is divided discontinuously.


In \cite{nuno}, the authors observed that landslides occurred  mostly  in  countries with a small number of electors in the 2008 US presidential election.
This indicates that    herders refer to local information,  and not global information.
In a previous study, we analyzed the case 
wherein  herders could see the $r$ previous  votes.
This  indicates that   voters can share information locally.
We discussed only  information cascade transition and could not discuss
 transition for super and normal diffusions.
We consider  
stochastic  differential  equations, which are strong tools, and use them for analysis  in this study. 
The case of several candidates remains to be investigated. 
In this study,  we investigated the case of  only two candidates\footnote{We have previously  discussed the cases of  several candidates  when all voters are analog herders \cite{Mori}.  The probability functions of the share of votes of the candidates obey  the gamma distributions. }
In our future studies, we shall consider these cases.

\appendix\section*{Acknowledgment}
This work was supported by
Grant-in-Aid for Challenging Exploratory Research, No. 21654054 (SM).

\def\thesection{Appendix \Alph{section}}
\section{Derivation of model  from Bayes' theorem}

We estimate the probability that  the candidate $C_{1}$ is correct
by using the previous votes.
We assume that the $(t+1)$th  voter is a herder.
 The voter estimates the posterior distribution
that $C_1$ is the correct candidate
 by using Bayes' theorem.
The voter can see $t$ votes.

Herders estimate the probability that $C_i$, where  $i=0,1$,  is   the correct candidate by using the previous votes.
$Pr(C_i)$ is the probability that the voter estimates  that $C_i$  is   the correct candidate. 
We set the prior distribution  as
$Pr(C_0)=Pr(C_1)=1/2$.
Here, we assume that the  herders believe  that all voters are independent and estimate the percentage of correct answers as  $\hat{q}$.
The  posterior  distribution is
\begin{eqnarray}
Pr(C_1|c_1=k)&=&\frac{1}{2}\frac{t!}{k!(t-k)!}\hat{q}^k(1-\hat{q})^{t-k},
\nonumber \\
Pr(C_0| c_1=k)&=&\frac{1}{2}\frac{t!}{k!(t-k)!}(1-\hat{q})^k\hat{q}^{t-k},
\end{eqnarray}
where  $c_1=k$ indicates that the number of  votes for $C_1$ is $k$ before the voter votes.
We can obtain
\begin{equation}
\frac{Pr(C_1|c_1=k)}{Pr(C_0|c_1=k)}=(\frac{\hat{q}}{1-\hat{q}})^{2k-t}=e^{{2\lambda}{(k-\frac{t}{2})/t}},
\end{equation}
where $\lambda= t \log\frac{\hat{q}}{1-\hat{q}}$.
Here, $\lambda$ increases as $t$ increases.
Hence, the herder can calculate the probability that the candidate $C_1$ is correct when the number of  votes for $C_1$  is $k$: 
\begin{equation}
Pr(C_1|c_1=k)=\frac{1}{2} 
[ \tanh \lambda (\frac{k}{t}-\frac{1}{2} )+1 ].
\end{equation}
In the  region  $t>>1$, $\lambda\rightarrow \infty $,  
a voter believes the public perception,  and 
the behavior of herders becomes similar to that of   digital herders without $\hat{q}=1/2$.
When $\hat{q}=1/2$, the herders estimate that  the previous votes  are  not useful to  estimate   the correct answer and $\lambda=0$.

\section{Derivation of stochastic differential equation} 

We use  $\delta X_\tau=X_{\tau+\epsilon}-X_\tau$ and $\zeta_\tau$, a standard i.i.d. Gaussian sequence; our objective is to identify the drift $f_\tau$ and  the variance $g^2_\tau$  such that
\begin{equation}
\delta X_\tau=f_\tau(X_\tau)\epsilon+\sqrt{\epsilon}g_\tau(X_\tau)\zeta_{\tau+\epsilon}.
\end{equation}
Given $X_\tau=x$, using the transition probabilities of $\Delta_n$, we get
\begin{eqnarray}
\textrm{E}(\delta X_\tau)&=&\epsilon \textrm{E}(\Delta_{[\tau/\epsilon]+1}-\Delta_{[\tau/\epsilon]})=\epsilon(2p_{[\frac{l/\epsilon+\tau/\epsilon}{2}],\tau/\epsilon}-1)
\nonumber \\
&=&
\epsilon[(1-p)(2q-1)+p 
\tanh  (\frac{\lambda x}{2\tau})].
\end{eqnarray}
Then, the drift term is $f_\tau(x)=(1-p)(2q-1)+p 
\tanh  (\lambda x/2\tau)$.
Moreover, 
\begin{equation}
\sigma^2(\delta X_\tau)=\epsilon^2
[
1^2p_{[\frac{l/\epsilon+\tau/\epsilon}{2}],\tau/\epsilon}
+(-1)^2(1-p_{[\frac{l/\epsilon+\tau/\epsilon}{2}],\tau/\epsilon})]
=\epsilon^2,
\end{equation}
  such  that 
$g_{\epsilon,\tau}(x)=\sqrt{\epsilon}.$
We can   obtain  $X_\tau$ such that it  obeys a diffusion equation with small additive noise:
\begin{equation}
\textrm{d}X_\tau=[(1-p)(2q-1)+p 
\tanh  (\frac{\lambda X_\tau}{2\tau})]\textrm{d}\tau+\sqrt{\epsilon}.
\label{ito}
\end{equation}

\section{Behavior of solutions   of  stochastic differential equation}

We  consider the stochastic differential equation
\begin{equation}
\textrm{d}x_\tau=
(\frac{l x_\tau}{\tau})\textrm{d}\tau+\sqrt{\epsilon},
\label{ito}
\end{equation}
where $\tau\geq1$.
If we set $l=p\lambda(1-\bar{v}^2)/2$, (\ref{ito}) is identical to   (\ref{ito4}).
Let $\sigma^2_1$
be the variance of $x_1$.
If $x_1$ is Gaussian   $(x_1\sim\textrm{N}(x_1,\sigma^2_1))$ or deterministic $(x_1\sim\delta_{x1})$, the law of $x_\tau$ ensures that the  Gaussian is in  accordance with density
\begin{equation}
p_\tau(x)\sim
\frac{1}{\sqrt{2\pi}\sigma_\tau}\textrm{e}^{-(x-\mu_\tau)^2/2\sigma_\tau^2},
\end{equation}
where $\mu_\tau=\textrm{E}(x_\tau)$ is the expected value of $x_\tau$ and 
$\sigma^2_\tau\equiv \nu_\tau$ is its variance.
If $\Phi_\tau(\xi)=\log(\textrm{e}^{\textrm{i}\xi x_\tau})$
 is the logarithm of the characteristic function of the law of $x_\tau$, we have\begin{equation}
\partial_\tau
\Phi_\tau(\xi)
=\frac{l}{\tau}\xi\partial_\xi\Phi_\tau(\xi)
-\frac{\epsilon}{2}\xi^2,
\end{equation}
and
\begin{equation}
\Phi_\tau(\xi)=\textrm{i}\xi \mu_\tau-\frac{\xi^2}{2}\nu_\tau.
\end{equation}
Identifying
the real and imaginary parts of $\Phi_\tau(\xi)$, we 
obtain the dynamics of  $\mu_\tau$ as
\begin{equation}
\dot{\mu}_\tau=\frac{l}{\tau}\mu_\tau.
\end{equation}
The solution for $\mu_\tau$ is
\begin{equation}
\mu_\tau=x_1\tau^l.
\end{equation}
The dynamics of $\nu_\tau$ are  given by the Riccati equation
\begin{equation}
\dot{\nu}_\tau=\frac{2l}{\tau}\nu_\tau+\epsilon.
\label{gp}
\end{equation}
If $\nu\neq 1/2$, we get
\begin{equation}
\nu_\tau=
\nu_1\tau^{2l}+\frac{\epsilon}{1-2l}(\tau-\tau^{2l}).
\end{equation}
If $l=1/2$, we get
\begin{equation}
\nu_\tau=\nu_1\tau+\epsilon\tau\textrm{log}\tau.
\end{equation}
We can  summarize the temporal behavior of the variance as 
\begin{equation}
\nu_\tau\sim\frac{\epsilon}{1-2l}\tau\hspace{1cm}\textrm{if}\hspace{0.5cm}l<\frac{1}{2},
\end{equation}
\begin{equation}
\nu_\tau\sim(\nu_1+\frac{\epsilon}{2l-1})\tau^{2l}\hspace{1cm}\textrm{if}\hspace{0.5cm}l>\frac{1}{2},
\end{equation}
\begin{equation}
\nu_\tau\sim\epsilon\tau\textrm{log}(\tau)\hspace{1cm}\textrm{if}\hspace{0.5cm}l=\frac{1}{2}.
\end{equation}


This model has three phases. If $l>1/2$ or $l=1/2$,
    $x_\tau/\tau$ converges  slower than  in  a binomial distribution.
These phases  are the super diffusion phases.  
If  $0<p<1/2$, $x_\tau/\tau$   converges as    in a binomial distribution.
This is the normal phase \cite{Hisakado2}.

\end{document}